\documentclass[RNAAS]{aastex631}

\usepackage{verbatim, amsmath, amsfonts, amssymb,amssymb}
\usepackage{multirow}
\usepackage[utf8]{inputenc}
\usepackage{empheq}
\usepackage[skins,theorems]{tcolorbox}
\usepackage{latexsym}
\usepackage{bigints}
\usepackage{natbib}
\usepackage{color}
\usepackage{tikz}

\shorttitle{AASTeX v6.3.1 Sample article}
\shortauthors{Gondhalekar et al.}
\graphicspath{{./}{figures/}}

\begin{document}

\title{\texttt{galmask}: A Python package for unsupervised galaxy masking}

\correspondingauthor{Rafael S. de Souza}
\email{drsouza@shao.ac.cn}
\correspondingauthor{Yash Gondhalekar}
\email{yashgondhalekar567@gmail.com}

\author[0000-0002-6646-4225]{Yash Gondhalekar}
\affiliation{Birla Institute of Technology and Science, Pilani, Goa, India}

\author[0000-0001-7207-4584]{Rafael S. de Souza}
\affiliation{Shanghai Astronomical Observatory,
Chinese Academy of Sciences, 80 Nandan Rd., Shanghai 200030, China}

\author[0000-0003-3220-0165]{Ana L. Chies-Santos}
\affiliation{Instituto de Física, Universidade Federal do Rio Grande do Sul (UFRGS), Av. Bento Gonçalves, 9500, Porto Alegre, RS, Brazil}
\affiliation{Shanghai Astronomical Observatory,
Chinese Academy of Sciences, 80 Nandan Rd., Shanghai 200030, China}


\begin{abstract}
Galaxy morphological classification is a fundamental aspect of galaxy formation and evolution studies. Various machine learning tools have been developed for automated pipeline analysis of large-scale surveys, enabling a fast search for objects of interest. However, crowded regions in the image may pose a challenge as they can lead to bias in the learning algorithm. 
In this Research Note, we present  \texttt{galmask},  an open-source package for unsupervised galaxy masking to isolate the central object of interest in the image. \texttt{galmask} is written in Python and can be installed from PyPI via the \texttt{pip} command. 
\end{abstract}



\section{Introduction} 

A galaxy's morphology encodes valuable information about the underlying physical processes driving their formation and evolution \citep{van2008,2013ApJ...774...47L, Conselice2014,2017AcASn..58....9F}. It correlates with physical properties, such as environmental density, merger history, and star formation rate. The first step in deriving the morphology of a galaxy in large-scale surveys and crowded regions is to isolate it from sources in the same field. Thus, there is considerable interest in generating automated astronomical source detection and segmentation tools \citep{1996A&AS..117..393B, 2015ApJS..220....1A} scalable to large-scale surveys. Examples include \texttt{Morpheus} \citep{2020ApJS..248...20H},  a deep learning approach for pixel-level classification using semantic segmentation that requires training the model with user-specified segmentation maps, and \texttt{galclean}  \citep{2018MNRAS.473.2701D}, which was designed to remove bright sources around a central galaxy by generating a non-target segmentation map and replacing the non-galaxy regions with a median background estimate.
Also, \citet{2020A&C....3300420F} developed an automated machine learning pipeline to perform detection, segmentation, and morphological classification of galaxies. 
A traditional machine learning classification scheme is usually designed to provide a single class per image. The learning method might be affected by neighboring objects such as stars and other galaxies in the same field. \texttt{galmask} is a general-purpose package to isolate the central object and remove unwanted detections. The only assumption is that the object of interest is placed near the center of the image.

\section{Data}

To showcase the capabilities of our method, we apply it to the
crowded field around the galaxy SDSS J095734.63+033901.7 taken with the Hyper Suprime-Cam (HSC) on the 8.2-m Subaru Telescope. We downloaded stamps in the $g$, $r$ and $i$- bands in fits format from hscMap\footnote{\url{https://hsc-release.mtk.nao.ac.jp/hscMap-pdr3/}}. The galaxy is located in the COSMOS field and the region of the CAMIRA cluster HSCJ095728+033956 \citep{oguri2018} at $z\sim0.16$. The equatorial coordinates of the galaxy are RA=09:57:34.6507 and DEC=+03:39:01.9612.

\clearpage
\section{Results and Conclusions}

The default configuration of \texttt{galmask} employs three main steps for the galaxy masking process. Firstly, it receives the original galaxy image, and convolves it with either a user-specified kernel or a normalized 2D Gaussian kernel with full width at half maximum, $\rm FWHM = 3$. Then it estimates an initial segmentation map using the \texttt{photutils} \citep{Bradley} library by selecting sources above a user-specified sigma threshold level.

The second step is deblending on the segmentation map to remove outliers and delineate the region around the central galaxy. For this, we use the \texttt{deblend\_sources} method from \texttt{photutils} that uses multi-thresholding and watershed segmentation to deblend nearby or overlapping sources. We empirically found deblending unnecessary in simple cases, and hence we keep this step optional. Local peaks, i.e., local maxima, are searched in each distinct labeled region of the partially-cleaned segmentation map from previous steps using the \texttt{peak\_local\_max} method from the \texttt{scikit-image} library \citep{scikit-image}.

The third and final step employs connected-component labeling \citep[CCL;][]{3918, Dillencourt92ageneral} to isolate the central object. CCL is an algorithm to label each connected component in an image. Let $S$ be a subset region of the original image. $S$ is a connected component if, for each pixel in $S$, there is a path to any other pixel in $S$ consisting of pixels that only belong to $S$. Our CCL implementation makes use of the \texttt{opencv-python} \citep{opencv_library} library.  We also ensure that the central galaxy is not masked during this step if the background sources are area-wise larger than the central galaxy or if they dominate the image region. Finally, we replace all non-source pixels with zero values in the final output image.

\begin{figure}[ht]
\includegraphics[width=1.\linewidth]{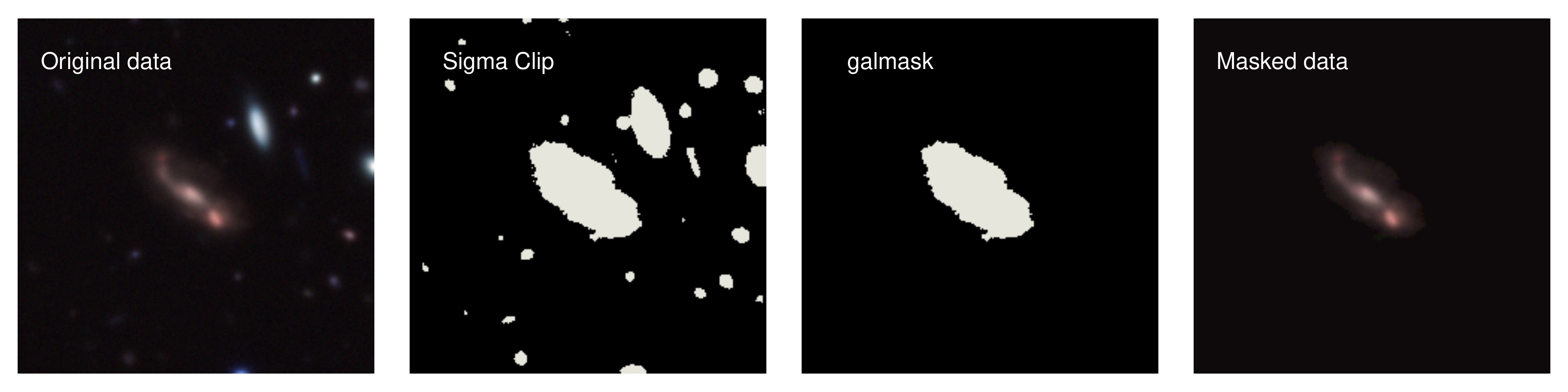}
\caption{Application of \texttt{galmask} on the crowded field around the galaxy SDSS J095734.63+033901.7 observed by Subaru/HSC.
The leftmost panel shows a $gri$ color composite image of the original field. The second panel shows the initial segmentation map computed by \texttt{galmask}. The third panel shows the final mask, and the fourth panel shows a color composite image of the masked galaxy. The map between the $gri$ bands and  RGB colors is made using an $\rm asinh$ stretch \citep[see, e.g.][]{Lupton2004}.}
\label{fig:galmaskExample}
\end{figure}

\autoref{fig:galmaskExample} shows the application of \texttt{galmask} to the region of the CAMIRA cluster HSCJ095728+033956. The panels depict different steps of the analysis. The original input image is displayed on the leftmost panel. The first segmentation, using sigma clipping, is depicted in the second panel, and the final mask of the central objects via CLL appears in the third panel. Furthermore, the rightmost panel provides a visualization of the masked galaxy image. \texttt{galmask} successfully removes the background sources surrounding the central galaxy and nearby sources.

\texttt{galmask} is available as an open-source software package on GitHub\footnote{\url{https://github.com/Yash-10/galmask/}} and Zenodo.\footnote{\url{https://doi.org/10.5281/zenodo.6626668}} The package can be installed via the pip command.  \footnote{\url{https://pypi.org/project/galmask/}}

\software{{\tt Astropy} \citep{astropy:2013, astropy:2018}, {\tt photutils} \citep{Bradley}, {\tt OpenCV} \citep{opencv_library}, {\tt scikit-image} \citep{scikit-image}.}

\begin{acknowledgments}
RSS thanks the National Natural Science Foundation of China, grant E045191001. ACS thanks the Chinese Academy of Sciences (CAS) President's International Fellowship Initiative (PIFI) through grant E085201009 and the Conselho Nacional de Desenvolvimento Científico e Tecnológico (CNPq) through grant CNPq-314301/2021-6. We thank Carolina Queiroz de Abreu Silva for useful discussions. This research made use of the {\tt Photutils} and {\tt Astropy} packages for
detection and photometry of astronomical sources. 
\end{acknowledgments}

%

\vspace{5mm}

\bibliography{ref}{}

\begin{thebibliography}{}
\expandafter\ifx\csname natexlab\endcsname\relax\def\natexlab#1{#1}\fi
\providecommand{\url}[1]{\href{#1}{#1}}
\providecommand{\dodoi}[1]{doi:~\href{http://doi.org/#1}{\nolinkurl{#1}}}
\providecommand{\doeprint}[1]{\href{http://ascl.net/#1}{\nolinkurl{http://ascl.net/#1}}}
\providecommand{\doarXiv}[1]{\href{https://arxiv.org/abs/#1}{\nolinkurl{https://arxiv.org/abs/#1}}}

\bibitem[{{Akhlaghi} \& {Ichikawa}(2015)}]{2015ApJS..220....1A}
{Akhlaghi}, M., \& {Ichikawa}, T. 2015, \apjs, 220, 1,
  \dodoi{10.1088/0067-0049/220/1/1}

\bibitem[{{Astropy Collaboration} {et~al.}(2013){Astropy Collaboration},
  {Robitaille}, {Tollerud}, {Greenfield}, {Droettboom}, {Bray}, {Aldcroft},
  {Davis}, {Ginsburg}, {Price-Whelan}, {Kerzendorf}, {Conley}, {Crighton},
  {Barbary}, {Muna}, {Ferguson}, {Grollier}, {Parikh}, {Nair}, {Unther},
  {Deil}, {Woillez}, {Conseil}, {Kramer}, {Turner}, {Singer}, {Fox}, {Weaver},
  {Zabalza}, {Edwards}, {Azalee Bostroem}, {Burke}, {Casey}, {Crawford},
  {Dencheva}, {Ely}, {Jenness}, {Labrie}, {Lim}, {Pierfederici}, {Pontzen},
  {Ptak}, {Refsdal}, {Servillat}, \& {Streicher}}]{astropy:2013}
{Astropy Collaboration}, {Robitaille}, T.~P., {Tollerud}, E.~J., {et~al.} 2013,
  \aap, 558, A33, \dodoi{10.1051/0004-6361/201322068}

\bibitem[{{Astropy Collaboration} {et~al.}(2018){Astropy Collaboration},
  {Price-Whelan}, {Sip{\H{o}}cz}, {G{\"u}nther}, {Lim}, {Crawford}, {Conseil},
  {Shupe}, {Craig}, {Dencheva}, {Ginsburg}, {Vand erPlas}, {Bradley},
  {P{\'e}rez-Su{\'a}rez}, {de Val-Borro}, {Aldcroft}, {Cruz}, {Robitaille},
  {Tollerud}, {Ardelean}, {Babej}, {Bach}, {Bachetti}, {Bakanov}, {Bamford},
  {Barentsen}, {Barmby}, {Baumbach}, {Berry}, {Biscani}, {Boquien}, {Bostroem},
  {Bouma}, {Brammer}, {Bray}, {Breytenbach}, {Buddelmeijer}, {Burke},
  {Calderone}, {Cano Rodr{\'\i}guez}, {Cara}, {Cardoso}, {Cheedella}, {Copin},
  {Corrales}, {Crichton}, {D'Avella}, {Deil}, {Depagne}, {Dietrich}, {Donath},
  {Droettboom}, {Earl}, {Erben}, {Fabbro}, {Ferreira}, {Finethy}, {Fox},
  {Garrison}, {Gibbons}, {Goldstein}, {Gommers}, {Greco}, {Greenfield},
  {Groener}, {Grollier}, {Hagen}, {Hirst}, {Homeier}, {Horton}, {Hosseinzadeh},
  {Hu}, {Hunkeler}, {Ivezi{\'c}}, {Jain}, {Jenness}, {Kanarek}, {Kendrew},
  {Kern}, {Kerzendorf}, {Khvalko}, {King}, {Kirkby}, {Kulkarni}, {Kumar},
  {Lee}, {Lenz}, {Littlefair}, {Ma}, {Macleod}, {Mastropietro}, {McCully},
  {Montagnac}, {Morris}, {Mueller}, {Mumford}, {Muna}, {Murphy}, {Nelson},
  {Nguyen}, {Ninan}, {N{\"o}the}, {Ogaz}, {Oh}, {Parejko}, {Parley}, {Pascual},
  {Patil}, {Patil}, {Plunkett}, {Prochaska}, {Rastogi}, {Reddy Janga},
  {Sabater}, {Sakurikar}, {Seifert}, {Sherbert}, {Sherwood-Taylor}, {Shih},
  {Sick}, {Silbiger}, {Singanamalla}, {Singer}, {Sladen}, {Sooley},
  {Sornarajah}, {Streicher}, {Teuben}, {Thomas}, {Tremblay}, {Turner},
  {Terr{\'o}n}, {van Kerkwijk}, {de la Vega}, {Watkins}, {Weaver}, {Whitmore},
  {Woillez}, {Zabalza}, \& {Astropy Contributors}}]{astropy:2018}
{Astropy Collaboration}, {Price-Whelan}, A.~M., {Sip{\H{o}}cz}, B.~M., {et~al.}
  2018, \aj, 156, 123, \dodoi{10.3847/1538-3881/aabc4f}

\bibitem[{{Bertin} \& {Arnouts}(1996)}]{1996A&AS..117..393B}
{Bertin}, E., \& {Arnouts}, S. 1996, \aaps, 117, 393,
  \dodoi{10.1051/aas:1996164}

\bibitem[{{Bradley} {et~al.}(2016){Bradley}, {Sipocz}, {Robitaille},
  {Tollerud}, {Deil}, {Vin{\'\i}cius}, {Barbary}, {G{\"u}nther}, {Bostroem},
  {Droettboom}, {Bray}, {Bratholm}, {Pickering}, {Craig}, {Pascual}, {Greco},
  {Donath}, {Kerzendorf}, {Littlefair}, {Barentsen}, {D'Eugenio}, \&
  {Weaver}}]{Bradley}
{Bradley}, L., {Sipocz}, B., {Robitaille}, T., {et~al.} 2016, {Photutils:
  Photometry tools}, Astrophysics Source Code Library, record ascl:1609.011.
\newblock \doeprint{1609.011}

\bibitem[{Bradski(2000)}]{opencv_library}
Bradski, G. 2000, Dr. Dobb's Journal of Software Tools

\bibitem[{{Conselice}(2014)}]{Conselice2014}
{Conselice}, C.~J. 2014, \araa, 52, 291,
  \dodoi{10.1146/annurev-astro-081913-040037}

\bibitem[{{de Albernaz Ferreira} \& {Ferrari}(2018)}]{2018MNRAS.473.2701D}
{de Albernaz Ferreira}, L., \& {Ferrari}, F. 2018, \mnras, 473, 2701,
  \dodoi{10.1093/mnras/stx2266}

\bibitem[{Dillencourt \& Samet(1992)}]{Dillencourt92ageneral}
Dillencourt, M.~B., \& Samet, H. 1992, Journal of the ACM, 39, 253

\bibitem[{{Fang} {et~al.}(2017){Fang}, {Lin}, \& {Kong}}]{2017AcASn..58....9F}
{Fang}, G.~W., {Lin}, Z.~S., \& {Kong}, X. 2017, Acta Astronomica Sinica, 58, 9

\bibitem[{{Farias} {et~al.}(2020){Farias}, {Ortiz}, {Damke}, {Jaque Arancibia},
  \& {Solar}}]{2020A&C....3300420F}
{Farias}, H., {Ortiz}, D., {Damke}, G., {Jaque Arancibia}, M., \& {Solar}, M.
  2020, Astronomy and Computing, 33, 100420,
  \dodoi{10.1016/j.ascom.2020.100420}

\bibitem[{{Hausen} \& {Robertson}(2020)}]{2020ApJS..248...20H}
{Hausen}, R., \& {Robertson}, B.~E. 2020, \apjs, 248, 20,
  \dodoi{10.3847/1538-4365/ab8868}

\bibitem[{{Lee} {et~al.}(2013){Lee}, {Giavalisco}, {Williams}, {Guo}, {Lotz},
  {Van der Wel}, {Ferguson}, {Faber}, {Koekemoer}, {Grogin}, {Kocevski},
  {Conselice}, {Wuyts}, {Dekel}, {Kartaltepe}, \& {Bell}}]{2013ApJ...774...47L}
{Lee}, B., {Giavalisco}, M., {Williams}, C.~C., {et~al.} 2013, \apj, 774, 47,
  \dodoi{10.1088/0004-637X/774/1/47}

\bibitem[{{Lupton} {et~al.}(2004){Lupton}, {Blanton}, {Fekete}, {Hogg},
  {O'Mullane}, {Szalay}, \& {Wherry}}]{Lupton2004}
{Lupton}, R., {Blanton}, M.~R., {Fekete}, G., {et~al.} 2004, \pasp, 116, 133,
  \dodoi{10.1086/382245}

\bibitem[{{Oguri} {et~al.}(2018){Oguri}, {Lin}, {Lin}, {Nishizawa}, {More},
  {More}, {Hsieh}, {Medezinski}, {Miyatake}, {Jian}, {Lin}, {Takada}, {Okabe},
  {Speagle}, {Coupon}, {Leauthaud}, {Lupton}, {Miyazaki}, {Price}, {Tanaka},
  {Chiu}, {Komiyama}, {Okura}, {Tanaka}, \& {Usuda}}]{oguri2018}
{Oguri}, M., {Lin}, Y.-T., {Lin}, S.-C., {et~al.} 2018, \pasj, 70, S20,
  \dodoi{10.1093/pasj/psx042}

\bibitem[{Samet \& Tamminen(1988)}]{3918}
Samet, H., \& Tamminen, M. 1988, IEEE Transactions on Pattern Analysis and
  Machine Intelligence, 10, 579, \dodoi{10.1109/34.3918}

\bibitem[{van~der Walt {et~al.}(2014)van~der Walt, {S}ch\"onberger,
  {Nunez-Iglesias}, {B}oulogne, {W}arner, {Y}ager, {G}ouillart, {Y}u, \& the
  scikit-image contributors}]{scikit-image}
van~der Walt, S., {S}ch\"onberger, J.~L., {Nunez-Iglesias}, J., {et~al.} 2014,
  PeerJ, 2, e453, \dodoi{10.7717/peerj.453}

\bibitem[{{van der Wel}(2008)}]{van2008}
{van der Wel}, A. 2008, \apjl, 675, L13, \dodoi{10.1086/529432}

\end{thebibliography}
\bibliographystyle{aasjournal}



\end{document}